\documentclass[letterpaper, 10 pt, conference]{ieeeconf}

 \IEEEoverridecommandlockouts
\usepackage{amsmath}
\usepackage{amsfonts}

\usepackage{tikz}
\usetikzlibrary{shapes.geometric, arrows}
\usetikzlibrary{positioning}
\tikzstyle{ad} = [rectangle, rounded corners, minimum width=3cm, minimum height=1cm,text centered, draw=black, fill=blue!30]
\tikzstyle{ag} = [rectangle, rounded corners, minimum width=2cm, minimum height=1cm,text centered, draw=black, fill=green!30]
\tikzstyle{cd} = [rectangle, minimum width=3cm, minimum height=1cm, text centered, draw=black, fill=orange!30]

\usepackage{tikz-network}

\input{mysymbol.sty}

 \newtheorem{theorem}{Theorem}
 
 \newtheorem{lemma}{Lemma}
 \newtheorem{proposition}{Proposition}
\newtheorem{example}{Example}

\usepackage{cite}
\usepackage{amsmath,amssymb,amsfonts}
\usepackage{algorithmic}
\usepackage{textcomp}
\usepackage{xcolor}
\def\BibTeX{{\rm B\kern-.05em{\sc i\kern-.025em b}\kern-.08em
    T\kern-.1667em\lower.7ex\hbox{E}\kern-.125emX}}
    
   \usepackage{graphicx}
   \graphicspath{{plots/pdf/}}
\begin{document}

\title{Information Preferences of Individual Agents in Linear-Quadratic-Gaussian Network Games}




\author{Furkan Sezer, \IEEEmembership{Student Member, IEEE}, Ceyhun Eksin, \IEEEmembership{Member, IEEE}
\thanks{Authors are with the Department of Industrial and Systems Engineering, Texas A\&M University, College Station, TX 77843 USA (e-mails: furkan.sezer@tamu.edu, eksinc@tamu.edu). This work was supported by NSF CCF-2008855.}}

\maketitle

\begin{abstract}

We consider linear-quadratic-Gaussian (LQG) network games in which agents have quadratic payoffs that depend on their individual and neighbors' actions, and an unknown payoff-relevant state. An information designer determines the fidelity of information revealed to the agents about the payoff state to maximize the social welfare. Prior results show that full information disclosure is optimal under certain assumptions on the payoffs, i.e., it is beneficial for the average individual. In this paper, we provide conditions based on the strength of the dependence of payoffs on neighbors' actions, i.e., competition, under which a rational agent is expected to benefit, i.e., receive higher payoffs, from full information disclosure. We find that all agents benefit from information disclosure for the star network structure when the game is symmetric and submodular or supermodular. We also identify that the central agent benefits more than a peripheral agent from full information disclosure unless the competition is strong and the number of peripheral agents is small enough. 
Despite the fact that all agents expect to benefit from information disclosure ex-ante, a central agent can be worse-off from information disclosure in many realizations of the payoff state under strong competition, indicating that a risk-averse central agent can prefer uninformative signals ex-ante.
\end{abstract}


\section{Introduction}

In an incomplete information network game, multiple players compete to maximize their individual payoffs that depend on the action of the player, the neighboring players' actions, and on unknown states. Incomplete information games are employed to model traffic flow in communication or transportation networks \cite{phan2012congestion,brown2017studies,wu2021value}, power allocation of users in wireless networks with unknown channel gains \cite{altman2003s,bacci2011pre}, oligopoly price competition \cite{molavi2016learning}, and coordination of autonomous teams \cite{eksin2013learning,eksin2014bayesian}. The information design problem refers to the determination of the information fidelity of the signals given to the players about the payoff state so that the induced actions of players maximize a system level objective. 

\begin{figure}
\centering
\begin{tikzpicture}[scale=0.85]
\footnotesize
\Vertex[label=Agent 1  $a_{1} (\omega_{1})$ ,size=2]{A} 
\Vertex[label=Agent 2  $a_{2} (\omega_{2})$ ,size=2, x=3,y=-2]{B}
\Vertex[label=Agent 3  $a_{3}( \omega_{3})$ ,size=2, x=-3,y=-2]{C}
\Vertex[label=Agent 4  $a_{4} (\omega_{4})$ ,size=2, x=3,y=2]{D}
\Vertex[label=Agent 5  $a_{5} (\omega_{5})$ ,size=2, x=-3,y=2]{E}
\Vertex[label=Inf. Designer $\zeta(\omega| \gamma)$,size=2.5, y=4, color=yellow ]{F}
\Edge(A)(B)
\Edge(A)(C)
\Edge(A)(D)
\Edge(A)(E)
\Edge[label=$\omega_{1}$,color=red](A)(F)
\Edge[label=$\omega_{2}$,color=red](F)(B)
\Edge[label=$\omega_{3}$,color=red](F)(C)
\Edge[label=$\omega_{4}$,color=red](F)(D)
\Edge[label=$\omega_{5}$,color=red](F)(E)
\end{tikzpicture}
\caption{Agents play a network game with individual payoffs that depend on their neighbors' actions and an unknown payoff state $\gamma$. An information designer sends a signal $w_{i}$ to each agent drawn from information structure $\zeta(\omega| \gamma)$. Agent $i$ takes an equilibrium action $a_{i}$ based on the received signal $\omega_{i}$ to maximize its expected utility. }
\vspace{-12pt}
\label{inf_design_diagram}
\end{figure}
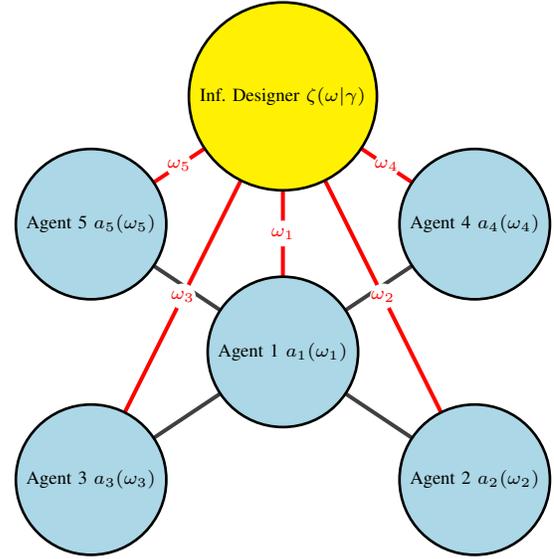

An information designer is an entity that is more informed about the realized payoff state than the players. Information designer selects an optimal probability distribution of signals based on the state realization with respect to its objective (see Fig. \ref{inf_design_diagram}). Various entities, e.g., a system operator overseeing the spectrum allocation, a market-maker, an independent system operator in the power grid, or the federal reserve, can be considered as an information designer. Information designers may define different objectives such as maximizing social welfare\cite{sezer2021social}, minimizing misinformation\cite{candogan2020misinformation}, or maximizing auction revenue\cite{emek2014}. In the absence of a (real) designer, an information design problem can be considered as a way to quantify the sensitivity of a system level objective to the information available to the players \cite{bergemann2016bayes, Bergemann2019} 

In this paper, we focus on social welfare maximization via information design in linear-quadratic-Gaussian (LQG) network games (Fig. \ref{inf_design_diagram}). Social welfare is defined as the aggregate utility of the players. In an LQG game, the players have quadratic payoff functions, and the state and the signals (types) come from a Gaussian distribution \cite{radner1962}. Under certain assumptions for the quadratic payoff coefficients, the rational behavior, defined as the Bayesian Nash equilibrium (BNE), in LQG games is unique. In \cite{sezer2021social}, we show that full information disclosure is the optimal solution to social welfare maximization under public information structures and/or common payoff states (see Theorem \ref{thm:sub-super}). 

While full information disclosure may be optimal from the system perspective, here we analyze the effect of such information disclosure policy on the payoffs of individual agents and its dependence on the centrality of the agents. We identify sufficient conditions for the individual preference of informative signals based on the payoff coefficients prior to realization of the state (ex-ante) in Theorem \ref{thm_main}. We leverage this result, and identify that both central and peripheral agents in a star network structure prefer information disclosure ex-ante for symmetric submodular and supermodular LQG games (Proposition \ref{prop_star}). In computing the benefit of information disclosure to individual agents, we find that a peripheral agent can benefit more than the central agent under full information disclosure if competition is strong and number of agents is small (Proposition \ref{prop_benefit}). In sum, the incentives of the agents and the system designer are in congruence ex-ante given the conditions considered. 

We find that joint incentives of individual agents and the system designer can cease to exist ex-post, i.e., after the realization of the payoff state. In contrast to Proposition \ref{prop_star}, the central agent prefers no information disclosure ex-post if realization of the payoff state is lower than expected. In the context of Bertrand competition among firms in networked markets, these results imply that some agents may not benefit from information disclosure when the competition among firms is strong. Ex-post analysis is no use to an agent because agents do not observe the realized payoff state, but they observe signals generated by the information designer based on the realized state. Still, the ex-post analysis of incentives imply that a risk averse central agent can prefer uninformative signals ex-ante. These results extend prior knowledge on the information design problem \cite{Bergemann2019,Ui2020} by providing a characterization of the benefit of informative signals on players' payoffs and its dependence on centrality of the players in network games with incomplete information.

\section{Information Design in LQG Network Games} \label{sec_model}

\subsection{Information Design in Incomplete Information Games} \label{sec_inf_design_problem}

An incomplete information game $G$ involves a set of $n\in \naturals^+$ players belonging to the set $\ccalN:=\{1,...,n\}$, each of which selects actions $a_i \in A_i$ to maximize the expectation of its individual payoff function $u_i(a,\gamma)$ where $a \equiv (a_{i})_{i \in \ccalN} \in A $ and $\bbgamma \equiv (\gamma_{i})_{i \in \ccalN} \in \Gamma$ correspond to an action profile and an unknown payoff state, respectively. Agents form expectations about their payoffs based on their beliefs/types $\omega_i \in \Omega$ about the state. We represent the incomplete information game by the tuple $G:= \{\ccalN, A, \{u_i\}_{i\in \ccalN}, \{\omega_i\}_{i\in \ccalN}\}$. 

A strategy of player $i$ in an incomplete information game maps each possible value of its type  $\omega_i\in \Omega$ to an action, i.e., $s_{i}: \Omega \rightarrow A_{i}$. A strategy profile $s = (s_{i})_{i \in \ccalN}$ is a BNE with respect to an information structure (distribution function) $\zeta$, if it satisfies the following inequality
\begin{equation}\label{eq_bne}
E_{\zeta}[u_{i}(s_{i}(\omega_{i}), s_{-i},\bbgamma )|\omega_{i}] \geq E_{\zeta}[u_{i}(a_{i}', s_{-i},\bbgamma )|\omega_{i}], 
\end{equation}for all $ a_{i}' \in A_{i}, \omega_{i} \in \Omega, i \in \ccalN$ where $s_{-i}=  (s_{j}(\omega_{j}))_{j\neq i}$ denotes the equilibrium strategy of all the other players, and $E_{\zeta}$ is the expectation operator with respect to the distribution function $\zeta$ and a prior distribution on the payoff state.

An information designer aims to optimize the expected value of its objective function $f(a,\bbgamma)$, e.g., social welfare, by deciding on an information structure $\zeta$ belonging to the feasible space of probability distributions $Z$, i.e., 
\begin{equation}\label{eq:orig-obj}
\max_{\zeta \in Z} E_{\zeta}[f(s, \bbgamma)]
\end{equation}
where $s$ is a BNE strategy profile for the game $G$ under $\zeta$. 

The timeline for the information design problem is as follows---also see Fig. \ref{inf_design_diagram}. 
\begin{enumerate}
\item Designer selects $\zeta \in Z$ and notifies players.
\item Payoff state $\bbgamma$ is realized.
\item Players observe signals $\omega$ drawn from $\zeta(\omega |\bbgamma)$.  
\item Players act according to BNE given $\zeta$.
\end{enumerate}



The information design problem in \eqref{eq:orig-obj} is intractable in the general case. Here, we focus on LQG network games which yield to a tractable semi-definite program formulation of the information design problem for designers' objectives $f(\cdot)$ quadratic in strategies and state.

\subsection{Linear-Quadratic-Gaussian (LQG) Network Games}
A LQG game corresponds to an incomplete information game with quadratic payoff functions and Gaussian information structures. In a LQG game, each player $i \in \ccalN$ decides on his action $a_{i} \in \mathbb{R}$ according to a payoff function of the form given below,
\begin{equation}\label{utility}
u_{i}(a , \gamma ) = - H_{ii}a_{i}^{2} - 2 \sum_{j \neq i}H_{ij}a_{i}a_{j} + 2\gamma_{i}a_{i} +d_{i}(a_{-i}, \gamma)
\end{equation}
where $a \equiv (a_{i})_{i \in N} \in \mathbb{R}^{n} $ and $\boldsymbol{\gamma} \equiv (\gamma_{i})_{i \in \ccalN} \in \mathbb{R}^{n}$. The term  $d_{i}(a_{-i},\boldsymbol{\gamma})$ is an arbitrary function of the opponents' actions $a_{-i}\equiv (a_{j})_{j \neq i}$ and payoff state $\boldsymbol{\gamma}$. We assume the utility function is strictly concave, i.e., $H_{ii}>0$ for all $i\in\ccalN$. We collect the coefficients of the utility function in a matrix $H = [H_{ij}]_{n\times n}$.
Payoff state $\boldsymbol{\gamma}\in \mathbb{R}^n$ follows a Gaussian distribution, i.e., $\bbgamma \sim \psi(\boldsymbol{\mu}, \Sigma)$ where $\psi$ is a multivariate normal probability distribution with mean $\boldsymbol{\mu}\in \mathbb{R}^n$ and covariance matrix $\Sigma\in \mathbb{R}^{n\times n}$. 
%
Each player $i \in \ccalN$ receives a private signal $\omega_{i} \in \Omega_{i} \equiv \mathbb{R}^{m_{i}}$ for some $m_i\in \mathbb{N}^+$. We define the information structure of the game  $\zeta(\omega|\boldsymbol{\gamma})$ as the conditional distribution of $ \omega\equiv (\omega_{i})_{i \in N}$ given $\boldsymbol{\gamma}$.  We assume the joint distribution over the random variables $(\omega,\boldsymbol{\gamma})$ is Gaussian; thus, $\zeta$ is a Gaussian distribution. For a positive definite matrix $H$, there is an unique BNE strategy that is linear in private signals. Moreover, we can obtain the BNE strategy by solving a set of linear equations \cite{radner1962}. 

We consider network games where the nodes are the players $\ccalN$, and edges $\ccalE$ determine the payoff dependencies, i.e., if $(i,j)\notin \ccalE$ then $H_{ij}=0$, otherwise $H_{ij}\in \mathbb{R}$ for $(i,j)\in \ccalE$. Next, we provide an example.

\begin{example}[Bertrand Competition in Networked Markets]
Firms determine the price for their goods ($a_{i}$) facing a marginal cost of production ($\gamma_i$). 
Firms compete over markets that are connected \cite{Bimpikis2019}.
The demand is a function of the price of all the firms, $q_{i}=\vartheta-\varpi a_{i} + \varrho\sum_{j\neq i} a_{j}$ with positive constants $\vartheta$, $\varpi$ and $\varrho$. 
Firm $i$'s profit is its revenue $q_{i}a_{i}$ minus the cost of production $\gamma_{i}q_{i}$,
\begin{align}\label{eq_bertrand}
 u_{i}(a, \gamma) &= q_{i}a_{i}-\gamma_{i}q_{i}
\end{align}
Nodes of networks correspond to a firm in Bertrand competition. If two nodes share an edge, they compete over the same market. For a star network, the central node can be a multinational firm competing with local competitors (peripheral nodes).
\end{example}






\section{Social Welfare Maximization via Information Design}
Social welfare is the sum of individual utility functions, 
\begin{align}
f(a, \boldsymbol{\gamma}) &= \sum_{i=1}^{n} u_{i}(a,\boldsymbol{\gamma}).
\end{align}
For LQG games, the social welfare maximization objective is a quadratic function. When the objective function $f(\cdot)$ is quadratic, we can rewrite \eqref{eq:orig-obj} as a Frobenius inner product of the coefficient matrix $F$ and the covariance matrix of $(a,\gamma)$ denoted using $X$ (see \cite{sezer2021social} for details of the derivation),
\begin{equation}\label{eq:fro_obj}
\max_{X \in  P_{+}^{2n}} F\bullet X := \max_{X \in  P_{+}^{2n}} \begin{bmatrix}
F_{11} & F_{12}\\ F_{21} & F_{22}
\end{bmatrix} \bullet \begin{bmatrix}
var(a) & cov(a, \boldsymbol{\gamma}) \\cov( \boldsymbol{\gamma},a)&var(\boldsymbol{\gamma}) 
\end{bmatrix}
\end{equation}
where 
\begin{equation} \label{eq:pos_def}
F=\begin{bmatrix}
-H & I \\ I& \textit{O}
\end{bmatrix}\; {\text{,} } \qquad F \bullet X := \sum_{i=1}^{2n}\sum_{j=1}^{2n} F_{ij} X_{ij}, 
\end{equation}
and we use $P^{2n}_{+}$ to denote the set of all $2n \times 2n$ symmetric positive semi-definite matrices.

In \eqref{eq:fro_obj}, the variance of actions $var(a)$, and covariance matrix of the actions and the payoff state $cov(\boldsymbol{\gamma},a)$ make up the decision variable $X$. We assume $F_{22}$ is an $n$ by $n$ zero matrix $\textit{O}$ because the variance of the payoff state $var(\boldsymbol{\gamma})$ does not depend on the information structure $\zeta$.
Given a LQG game, following set of constraints induce the BNE (see Proposition 3, \cite{Ui2020}):
\begin{equation} \label{eq:action_constraints}
\sum_{j \in \ccalN}H_{ij} cov(a_{i},a_{j})=cov(a_{i},\gamma_{i}) \; \; \forall i\in \ccalN.
\end{equation}
Rewriting the above constraints using the covariance matrix $X$, we have the following design problem: 
\begin{align}
\max_{X\in  P_{+}^{2n}}\,& F\bullet X
\label{eq:mod1}\\
\text{s.t. }& \;M_{kl}\bullet X = cov(\gamma_{k},\gamma_{l}), \quad\forall k,l \in \ccalN \text{ with } k\leq l \label{eq:mod2} \\
&R_{k}\bullet X = 0 \quad \forall k \in \ccalN\label{eq:mod3} 
\end{align}
where $M_{kl}= [[M_{kl}]_{ij}]_{2n \times 2n} \in P^{2n}$ and $R_{k}= [[R_{k}]_{ij}]_{2n\times 2n} \in P^{2n}$ are defined as 
\begin{equation*}
[M_{kl}]_{ij}=\begin{cases}
1/2 \quad\text{ if } k<l, i=n+k, j=n+l\\
1/2 \quad\text{ if } k<l, i=n+l, j=n+k\\
1 \quad\text{ if } k=l, i=n+k, j=n+l\\
0 \quad \text{otherwise,}
\end{cases}  
\end{equation*}
\begin{equation*}
[R_{k}]_{ij}=\begin{cases}
H_{kk}&if \quad  i = j = k,  \\
H_{kj}/2  & if \quad  i = k, 1\leq j \leq n, j\neq k, \\
-1/2 &if \quad i = k, j= n + k,\\
H_{ki}/2 &if\quad  j = k, 1\leq i \leq n, i\neq k \\
-1/2 &if \quad j = k, i= n + k,\\
0 & \text{otherwise.}
\end{cases}
\end{equation*}
The equality constraints in \eqref{eq:mod2} assign the covariance matrix of the state $var(\boldsymbol{\gamma})$ to the corresponding elements of $X$. Constraints in \eqref{eq:mod3} correspond to the BNE conditions given in \eqref{eq:action_constraints}.
Note that the original information design problem \eqref{eq:orig-obj} is transformed to the maximization of a linear function of a positive semi-definite matrix $X$ subject to linear constraints, i.e., it is a SDP. 

Using the SDP formulation above, we have that {\it full information disclosure}, i.e., signals that reveal the payoff state, is an optimal strategy for the information designer under certain cases.

\begin{theorem}[Proposition 7-8,\cite{sezer2021social}]\label{thm:sub-super}
Full information disclosure is the optimal solution to \eqref{eq:mod1}-\eqref{eq:mod3} for social welfare maximization if either of the following conditions hold:
\begin{itemize}
    \item[{\bf (a)}] $H$ is positive definite and information designer reveals a single (public) signal.
    \item[{\bf (b)}] $H$ is symmetric and there is a common payoff state, i.e. $\gamma_{i}=\gamma_{j}, \, \forall i,j \in \ccalN$.
\end{itemize}
 \end{theorem}
 See \cite{sezer2021social} for the proof and details. 

 We interpret the results in Theorem \ref{thm:sub-super} for network games. Theorem \ref{thm:sub-super}(a) implies that if $H$ is diagonally dominant, then full information disclosure is optimal for public signal structures. In the context of Bertrand competition, this result implies that if every firm is to receive the same signal on the cost of their production, it is preferable to select the signal from a probability distribution which corresponds to full information disclosure. 
 According to Theorem \ref{thm:sub-super} (b), for network games with symmetric $H$, full information disclosure is optimal given a common payoff state. A common payoff state corresponds to a common marginal cost for firms in Bertrand competition. This result implies that each firm should receive a signal that reveals the designer's full information,  regardless of the values of the constants in $q_i$. 
%
%
It is possible that while full information is preferred to maximize the aggregate utility, some players in the network may be worse off by full information disclosure. Next, we analyze the ex-ante information structure preferences of individual agents based on their position in the network.

\section{Ex-ante Information Structure Preferences of Agents based on Network Position}\label{Sec_payoff} 
Equilibrium actions $a\in \mathbb{R}^{n}$ under full and no information disclosure are given by $H^{-1}\boldsymbol{\gamma}$ and $H^{-1}\boldsymbol{\mu}$ (see Lemma \ref{thm_public} in the appendix) when there is a common payoff state $\gamma$ i.e $\gamma_{i}=\gamma, \forall i \in \ccalN$ and public information structures. In this case, individual equilibrium actions under full and no information disclosure are given by $a_i=\gamma [H^{-1} \bbone]_{i}$ and $a_i=\mu[H^{-1} \bbone]_{i}$, respectively for $i\in \ccalN$ where $\bbone \in \mathbb{R}^n$ is a vector of ones and $[\cdot]_i$ represents the $i$th element of a vector. In this section, we treat the actions as random variables where we assume $\gamma \sim \psi(\mu, \sigma^{2})$ and $\mu \sim \psi(\mu_{0}, \sigma_{0}^{2})$.  




\begin{theorem} \label{thm_main}
Consider a LQG network game with common payoff state $\gamma$ and public information structures. If 
\begin{align}\label{eq_delta_util_ex_ante_final}
  [ H^{-1} \boldsymbol{1}]_{i} \bigg(2- H_{ii} [H^{-1} \boldsymbol{1}]_{i} - 2 \sum_{j \neq i}H_{ij} [H^{-1} \boldsymbol{1}]_{j} \bigg)>0, 
\end{align}
then full information disclosure is preferable by agent $i \in \ccalN$ over no information disclosure. 
\end{theorem}
\begin{proof}
If agent $i$'s expected utility given full information disclosure is larger than its expected utility at no information disclosure, then full information disclosure is preferable. We begin by computing agent $i$'s expected utility under full information disclosure by plugging in $a=\gamma H^{-1}\bbone$ into \eqref{utility}:
\begin{align}\label{eq_full_util_ex_ante}
E[u_{i}&(a, \gamma)] =E[\gamma^{2}] [H^{-1} \boldsymbol{1}]_{i} \bigg(2- H_{ii} [ H^{-1} \boldsymbol{1}]_{i} \nonumber \\ & - 2 \sum_{j \neq i}H_{ij}[  H^{-1} \boldsymbol{1}]_{j} \bigg)  +E[d_{i}(a_{-i}, \gamma)]
\end{align}
Next, we plug in  $a=\mu H^{-1}\bbone$ into \eqref{utility} for no information disclosure: 
\begin{align}\label{eq_no_util_ex_ante}
E[u_{i}&(a, \gamma)] =[ H^{-1} \boldsymbol{1}]_{i}  \bigg(E[\mu^{2}](- H_{ii}[H^{-1} \boldsymbol{1}]_{i} \nonumber \\ &- 2 \sum_{j \neq i}H_{ij}[H^{-1} \boldsymbol{1}]_{j}) + 2E[\gamma \mu]\bigg)  +E[d_{i}(a_{-i}, \gamma)]
\end{align}

We subtract \eqref{eq_no_util_ex_ante} from \eqref{eq_full_util_ex_ante}:
\begin{align}\label{eq_delta_util_ex_ante}
E[\Delta u_{i}(a, \gamma)] &= [ H^{-1} \boldsymbol{1}]_{i} \bigg (E[\gamma^{2}-\mu^{2}] ( - H_{ii}[ H^{-1} \boldsymbol{1}]_{i} \nonumber \\&- 2 \sum_{j \neq i}H_{ij}[ H^{-1} \boldsymbol{1}]_{j} ) + 2E[\gamma^{2}-\gamma\mu]\bigg)
\end{align}

Given that $\gamma \sim \psi(\mu, \sigma^{2})$ and $\mu \sim \psi(\mu_{0}, \sigma_{0}^{2})$, we have $E[\mu^2]-\mu_{0}^{2}=\sigma_{0}^{2}$, $E[\gamma^{2}]= \sigma^{2}+\sigma_{0}^{2}+\mu_{0}^{2}$, and $E[\gamma \mu] = \sigma_{0}^{2}+\mu_{0}^{2}$. We substitute these values in \eqref{eq_delta_util_ex_ante}, and simplify to get the condition in \eqref{eq_delta_util_ex_ante_final}.
\end{proof}

\subsection{Information Structure Preferences under Star Network}

We showcase Theorem \ref{thm_main} by applying to LQG games over a star network. A star network is comprised of a central agent ($i=1$) and $n-1$ peripheral agents ($j\in\ccalN \setminus \{1\}$). We derive information structure preference conditions for both the central agent and peripheral agents. In the following, we consider a payoff coefficients matrix where $H_{ii}=1$ and $H_{ij}=\beta$, for $(i,j) \in \ccalE$, and $\beta\in \mathbb{R}$.
\begin{proposition}\label{prop_star}
 If $(n-1)\left \vert \beta \right \vert <1$, then full information disclosure is  preferred over no information disclosure by all both the central agent and the peripheral agents in the star network. 
\end{proposition}
\begin{proof}
We compute $[H^{-1} \boldsymbol{1}]_{i}$, for $i \in \ccalN$:
\begin{equation}\label{eq_h_inverse}
[H^{-1} \boldsymbol{1}]_i = \frac{|\ccalN_i|\beta-1}{(n-1)\beta^{2}-1} \; \text{ for }\; i \in \ccalN
\end{equation}
where $\ccalN_i:\{j:(i,j)\in \ccalE\}$ denotes the neighbors of agent $i$, and $|\ccalN_i|$ denotes its cardinality. We check the condition in \eqref{eq_delta_util_ex_ante_final} for the central agent, say $i=1$, by substituting in \eqref{eq_h_inverse}, $|\ccalN_1|=n-1$ and $|\ccalN_j|=1$ for $j\in \ccalN \setminus \{1\}$,
\begin{align}\label{eq_delta_util_ex_ante_central1}
\frac{(n-1)\beta-1}{(n-1)\beta^{2}-1} \bigg( 2- \frac{(n-1)\beta-1 +2(n-1)\beta (\beta-1)}{(n-1)\beta^{2}-1} \bigg)> 0.
\end{align}
We simplify \eqref{eq_delta_util_ex_ante_central1} to get $((n-1)\beta-1)((n-1)\beta-3) >0$.
Solving the quadratic inequality, we get that no information disclosure could be preferable only if 
\begin{align}\label{eq_delta_util_ex_ante_central6}
 1<(n-1)\beta-1) <3.
\end{align}
Given that $(n-1)\beta<1$, full information disclosure is always preferable to no information disclosure by the central agent.  

Now we consider peripheral agents $j\in \ccalN \setminus \{1\}$. We check the condition \eqref{eq_delta_util_ex_ante_final} for a peripheral agent by substituting in \eqref{eq_h_inverse}, $|\ccalN_1| = n-1$, and $|\ccalN_j|=1$ for $j\in \ccalN\setminus \{1\}$:
\begin{align}\label{eq_delta_util_ex_ante_peripheral1}
  \frac{\beta-1}{(n-1)\beta^{2}-1} \bigg(2-   \frac{\beta(\beta-1) + 2((n-1)\beta-1)}{(n-1)\beta^{2}-1}  \bigg) >0.
\end{align}
Above equation simplifies to $(\beta-1)^{2} >0$ which is always satisfied. 
This means $E[\Delta u_{i}(a, \gamma)] $ is always positive. Therefore, full information disclosure is always preferable over no information disclosure by the peripheral agents.
\end{proof}
This result shows that all agents regardless of their position in the star network are expected to benefit from information disclosure. Next, we identify the region for $\beta$ where the expected benefit of the information disclosure to a peripheral agent is higher than that of the central agent.

\begin{figure}
\centering
\includegraphics[width=0.9\linewidth]{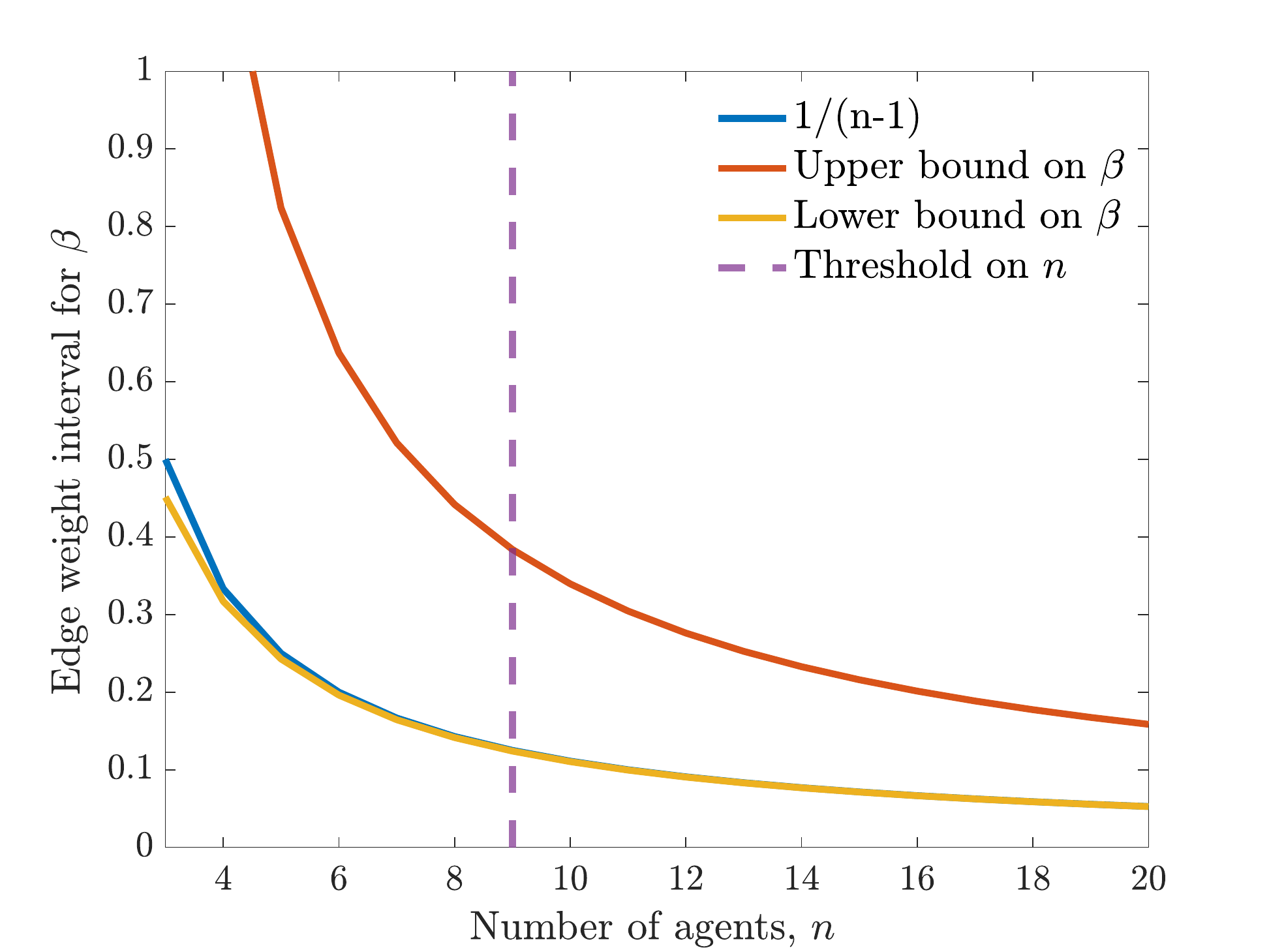}
\caption{We plot \eqref{eq_gain} for number of agents from 3 to 20. We also plot positive definiteness condition we impose on $\beta$, i.e., $(n-1)\beta < 1$. Indeed, the positive definiteness line ($1/(n-1)$) crosses below the lower bound in \eqref{eq_gain} at $n>9$, indicating that the central agent benefits more than a peripheral agent from information disclosure. }
\vspace{-12pt}
\label{fig_edge_weight} 
\end{figure}

\begin{proposition}\label{prop_benefit}
If 
\begin{equation}\label{eq_gain}
\frac{2(n-1)-\sqrt{\nu(n)}}{n(n-2)}< \beta < \frac{2(n-1)+\sqrt{\nu(n)}}{n(n-2)}
\end{equation}
where $\nu(n) =n^{2}-2n+4$, then the gain of a peripheral agent from information disclosure is higher than the gain of the central agent. For $\beta$ values outside  interval \eqref{eq_gain}, the gain of the central agent is higher than that of a peripheral agent.
\end{proposition}
\begin{proof}
We consider the difference between $E[\Delta u_{1}(a, \gamma)]$ in \eqref{eq_delta_util_ex_ante_central1} and $E[\Delta u_{j}(a, \gamma)]$ in \eqref{eq_delta_util_ex_ante_peripheral1} for $j \in \ccalN \setminus \{1\}$ to get 
\begin{align}\label{eq_diff1}
 E[&\Delta u_{1}(a, \gamma)] - E[\Delta u_{j}(a, \gamma)]=\nonumber \\&  \frac{((n-1)\beta-1)((n-1)\beta-3) - (\beta-1)^{2}}{((n-1)\beta^{2}-1)^{2}} >0.
\end{align}
We remove the positive valued denominator, and simplify the numerator to get
\begin{equation}
n(n-2)\beta^{2}-4(n-1)\beta+3 >0.\label{eq_diff2}
\end{equation}
Solving quadratic inequality \eqref{eq_diff2} indicates that when $\beta$ is in the range given in \eqref{eq_gain}, $E[\Delta u_{1}(a, \gamma)] - E[\Delta u_{j}(a, \gamma)] <0$. Thus, a peripheral agent benefits more than the central agent from full information disclosure. For $\beta$ values outside of the interval, we have  $E[\Delta u_{1}(a, \gamma)] - E[\Delta u_{j}(a, \gamma)] > 0$.
\end{proof}
In Fig. \ref{fig_edge_weight}, we plot the upper and lower bound values in \eqref{eq_gain} as a function of $n$. We observe the bounds get closer as $n$ increases. When we contrast these bounds with the bound for positive-definiteness, i.e., $\beta<1/(n-1)$, we observe that the upper bound is not realized for any $\beta$ value. For $n>9$, the positive definiteness condition implies that the lower bound cannot be exceeded. Thus, the central agent always benefits more than a peripheral agent for $n>9$. 


\begin{figure}
\centering
\begin{tabular}{c}
\includegraphics[width=0.9\linewidth]{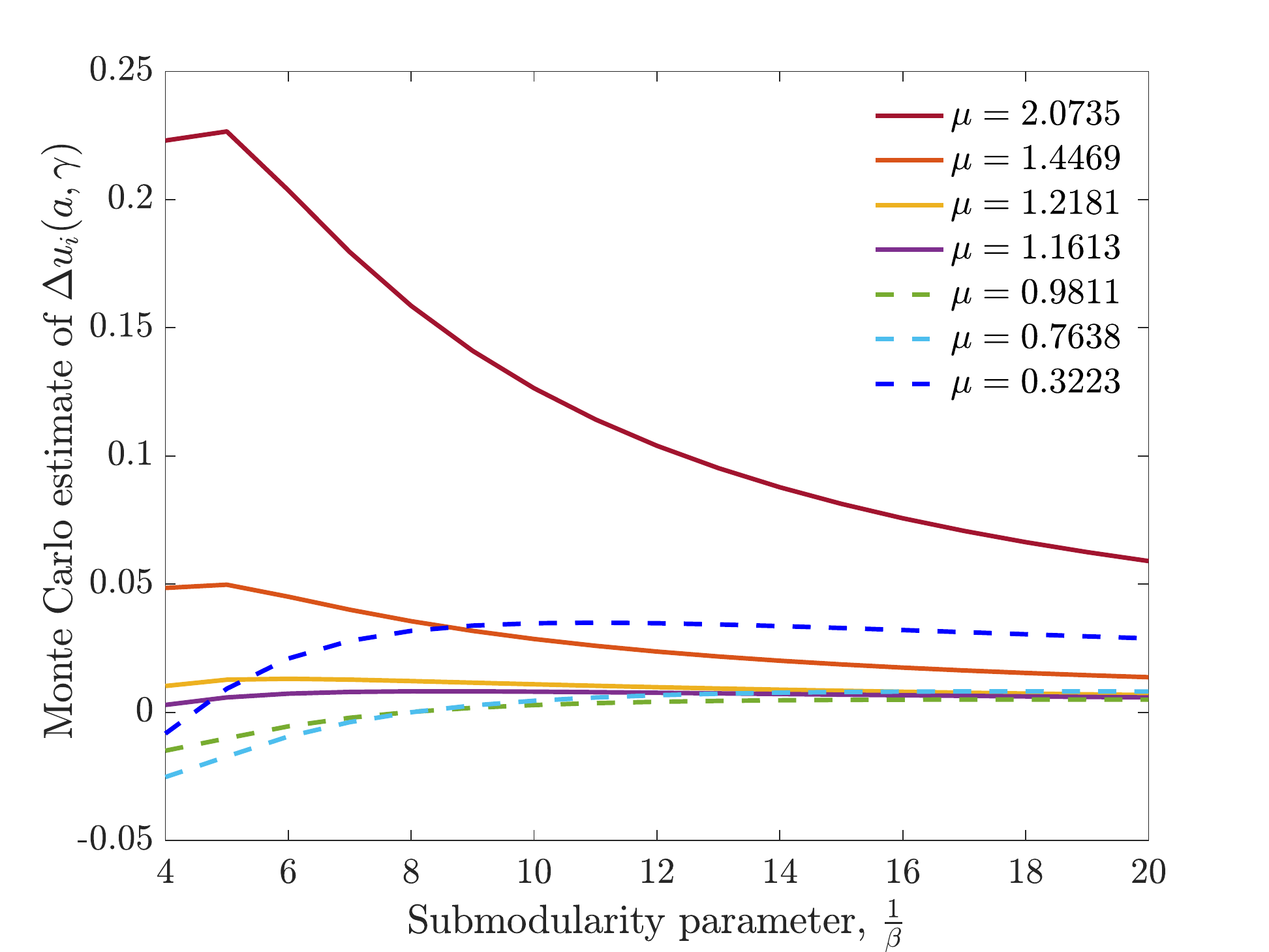}\\(a) Central agent\\
\includegraphics[width=0.9\linewidth]{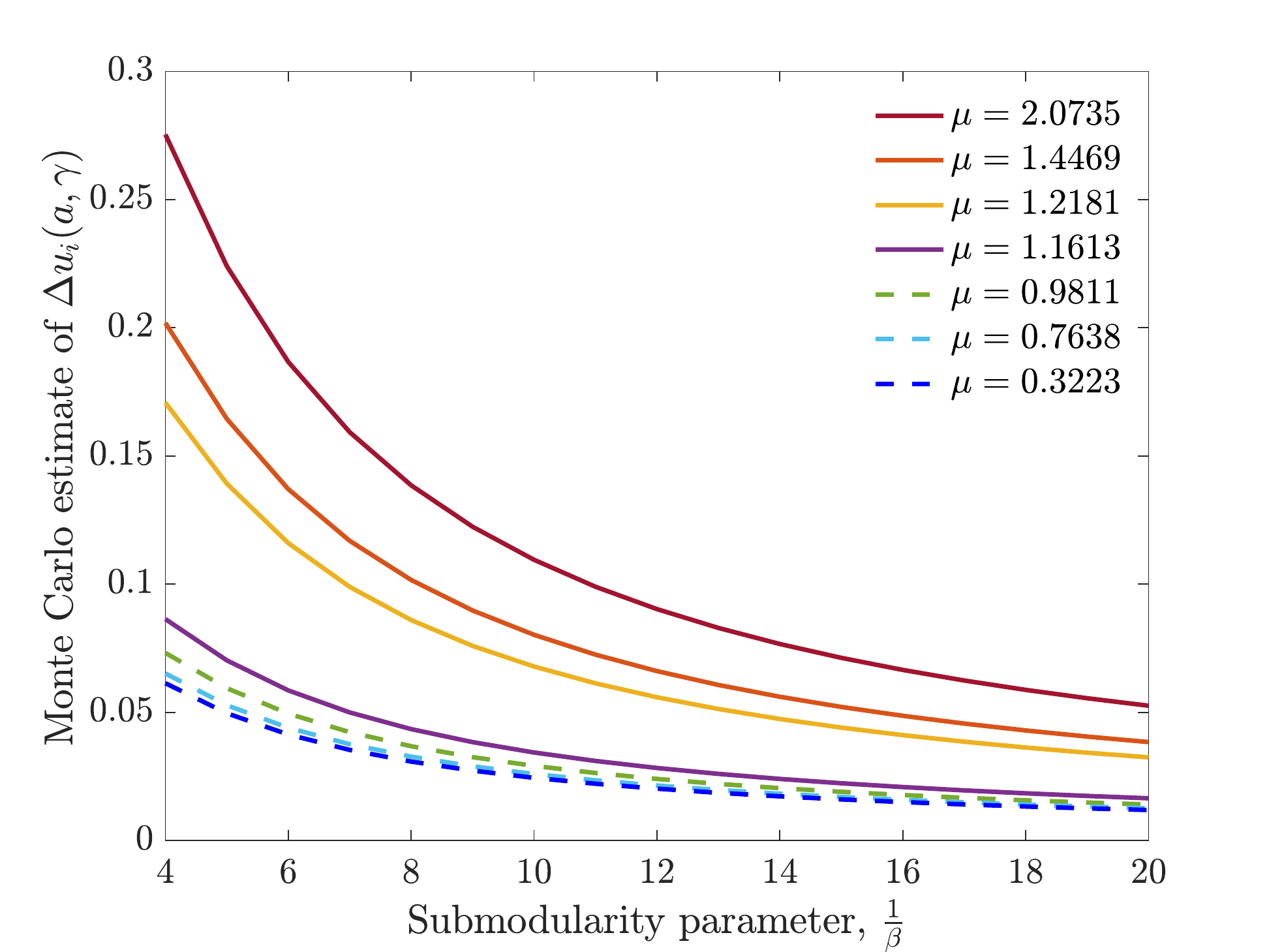}\\
(b) Peripheral agent
\end{tabular}
\caption{Ex-post information preference estimates of central and peripheral agents in submodular games on a star network with $n=4$. Lines show seven realized $\mu$ values generated from $\mu \sim \psi(\mu_0=1, 0.3^2)$. Dashed lines indicate $\mu<\mu_0$. Solid lines indicate $\mu>\mu_0$. For each $\mu$ and $\beta$ value, $1000$ $\gamma$ values are generated from $\psi(\mu, 0.1^2)$. We estimate $\delta u_i(a,\gamma)$ by averaging the values over $\gamma$ realizations. For large $\beta$ values, full information disclosure may not preferred by the central agent when $\mu<\mu_0$. }
\vspace{-12pt}
\label{fig_sub_line} 
\end{figure}

\section{Ex-post Information Structure Preferences \label{sec_num}}
Depending upon the realizations of $\mu$ and $\gamma,$ agents may want to prefer no information disclosure ex-post. 
We can say an agent prefers full information disclosure over no information disclosure if the change in utility function 
$\Delta u_{i}(a, \gamma) >0,$ and vice versa upon realization of $\mu$ and $\gamma$. We  express $\Delta u_{i}(a, \gamma)$ as follows by removing the expectation operator in  \eqref{eq_delta_util_ex_ante}, 
\begin{align}\label{eq_delta_util}
\Delta u_{i}(a, \gamma) &= (\gamma-\mu)[ H^{-1} \boldsymbol{1}]_{i} \bigg((\gamma+\mu) ( - H_{ii}[ H^{-1} \boldsymbol{1}]_{i}\nonumber \\&- 2 \sum_{j \neq i}H_{ij}[ H^{-1} \boldsymbol{1}]_{j}) + 2\gamma \bigg).
\end{align}
%
We estimate \eqref{eq_delta_util} numerically via Monte Carlo simulation for submodular ($\beta<0$) and supermodular ($\beta>0$) games. In submodular games, agents' actions are strategic substitutes, i.e., when agent $j$ increases its action agent $i$'s incentive to increase its action decreases ($\frac{\partial^{2} u_{i}}{\partial a_{i}\partial a_{j}} < 0$). In supermodular games, agents' actions complement each other, i.e., when agent $j$ increases its action agent $i$'s incentive to increase its action increases ($\frac{\partial^{2} u_{i}}{\partial a_{i}\partial a_{j}}>0$)---see Section 3, \cite{jackson2015games}. The Bertrand competition with payoffs in \eqref{eq_bertrand} is an example of a supermodular game. 

\begin{figure}[!htbp]
\centering
\begin{tabular}{c}
\includegraphics[width=0.85\linewidth]{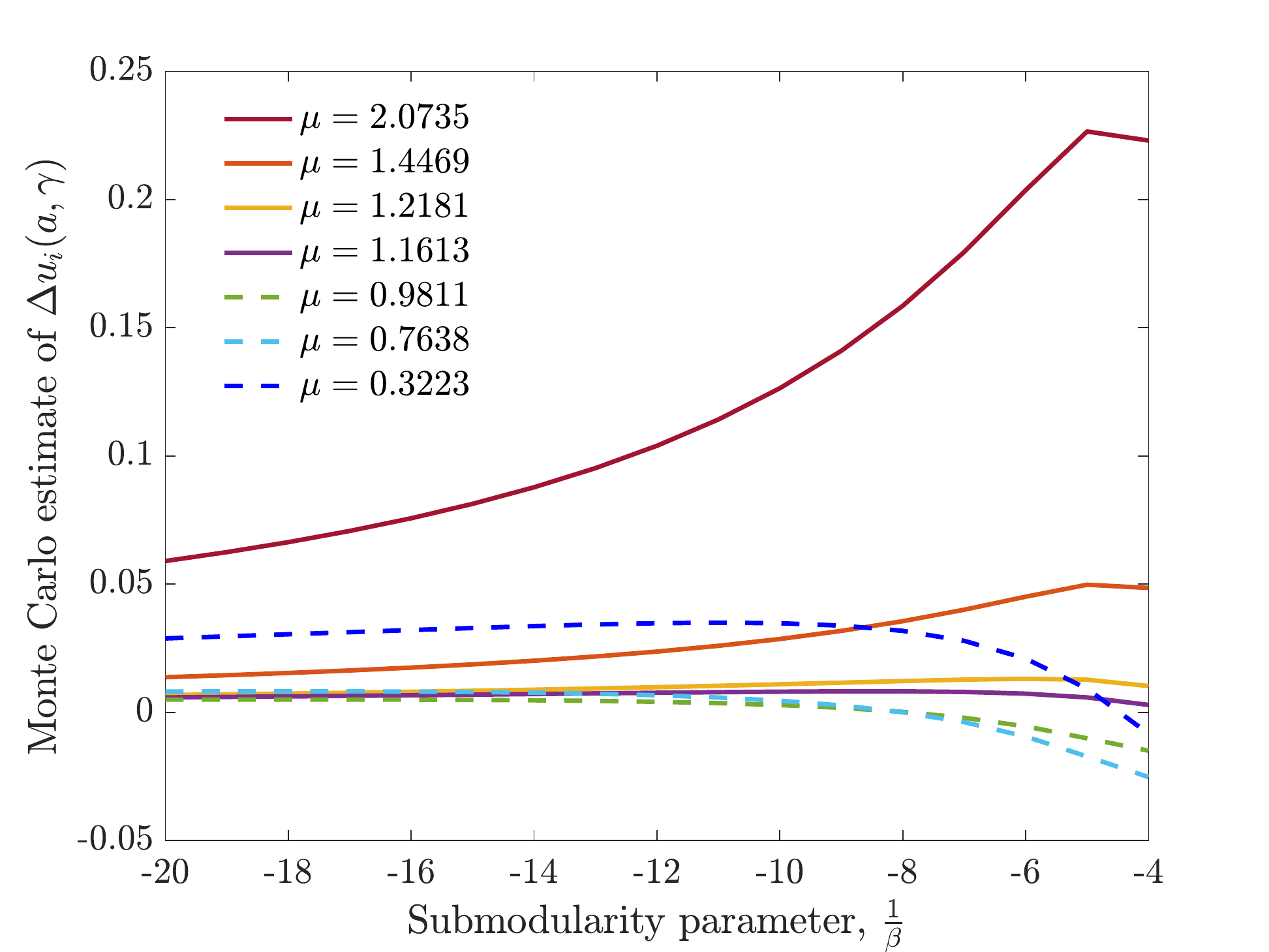}\\
(a) Central agent\\
\includegraphics[width=0.85\linewidth]{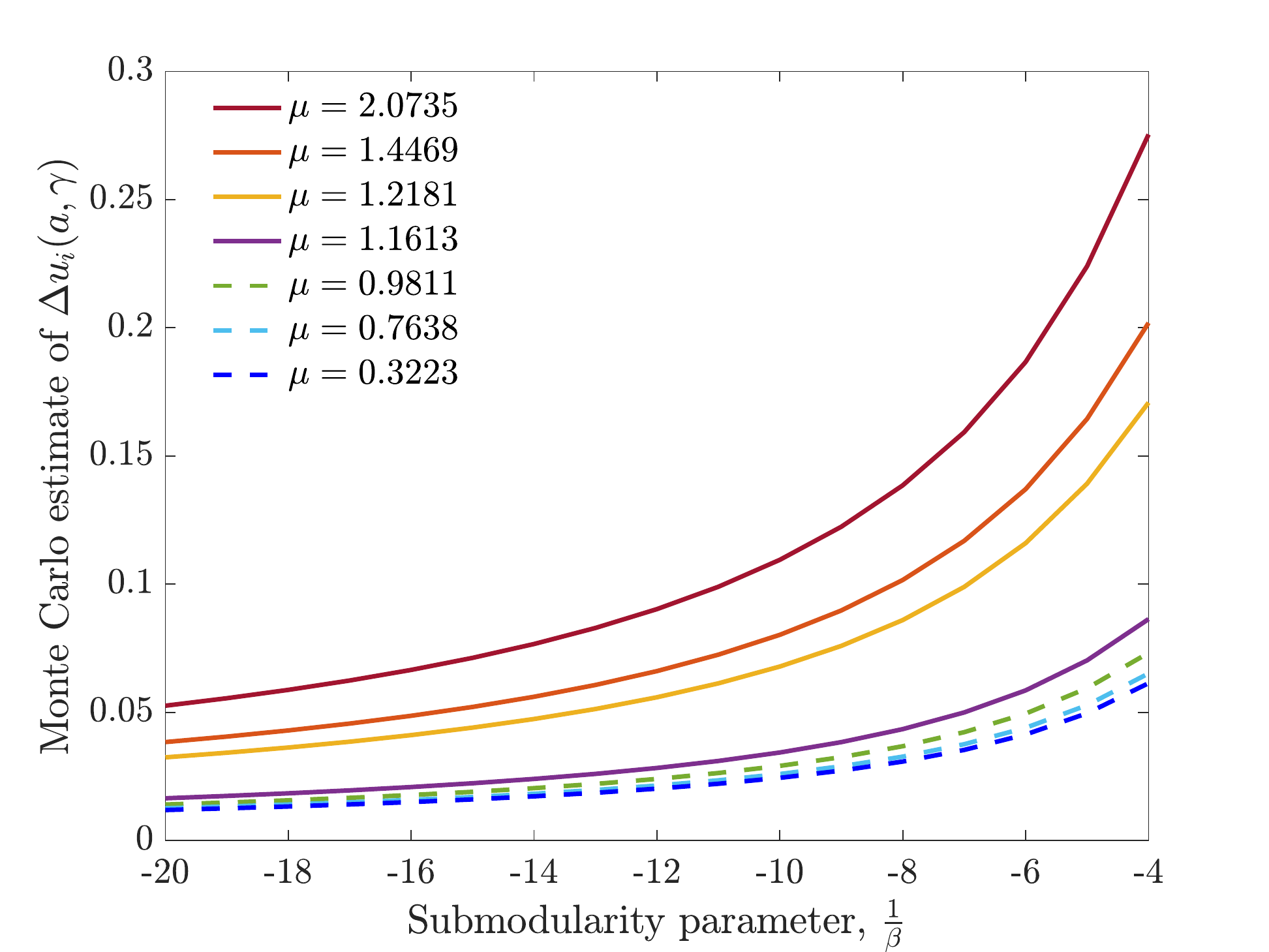}\\
 (b) Peripheral agent
\end{tabular}
\caption{Ex-post information preference estimates of central and peripheral agents in supermodular games on a star network with $n=4$. Lines show seven realized $\mu$ values generated from $\mu \sim \psi(\mu_0=1, 0.3^2)$. Dashed lines indicate $\mu<\mu_0$. Solid lines indicate $\mu>\mu_0$. For each $\mu$ and $\beta$ value, $1000$ $\gamma$ values are generated from $\psi(\mu, 0.1^2)$. We estimate $\delta u_i(a,\gamma)$ by averaging the values over $\gamma$ realizations. For large $|\beta|$ values, full information disclosure is not preferred by the central agent when $\mu<\mu_0$.
}\vspace{-12pt}
\label{fig_sup_line}
\end{figure}

We compute $\Delta u_{i}(a, \gamma)$ for submodular and supermodular games in Figs. \ref{fig_sub_line} and \ref{fig_sup_line}, respectively. In particular, we generate $\mu$ values from $\psi(\mu_0=1,0.3^{2})$, and $\gamma$ values from $\psi(\mu, 0.1^{2})$. We estimate $\Delta u_{i}(a, \gamma) $ for every combination of $\beta$ and $\mu$ value by averaging over realizations of $\gamma$.

In both types of games, the average change in utility function over realizations of $\mu$ is positive indicating that information disclosure is preferable and confirming Proposition \ref{prop_star}. The value of information decreases on average for both central and peripheral agents in both types of games as submodularity parameter $ \frac{1}{\left \vert \beta \right \vert}  $ increases. This is reasonable because the dependence of the payoffs on others' actions reduces as $\left \vert \beta \right \vert $ decreases. In both of the games, central agent prefers no information disclosure ex-post when realized $\mu$ is less than $\mu_{0}$ and the absolute value of submodularity parameter is low (Figs. \ref{fig_sub_line}(a) and \ref{fig_sup_line}(a)). Otherwise, the central agent prefers full information disclosure ex-post. This indicates a risk-averse central agent may prefer no information disclosure ex-ante. For instance, a multinational company in a Bertrand competition with local firms may prefer that information remains hidden when the production costs are high and competition is stiff. In contrast, a peripheral agent always prefers full information disclosure regardless of the realized $\mu$ values (Figs. \ref{fig_sub_line}(b) and \ref{fig_sup_line}(b)).

\section{Conclusion}
We considered whether the incentives of agents in a network game align with the information designer's objective to maximize social welfare or not. Given only prior information about the payoff state, agents in a star network preferred full information disclosure. Unless the competition is strong and the number of agents is small, the central agent benefited more than a peripheral agent from full information disclosure. 
However, ex-post incentive estimates showed that a risk-averse central agent can prefer no information disclosure ex-ante, while the peripheral agents continue to benefit from information disclosure. 


\bibliography{ref}
\bibliographystyle{ieeetr}

\appendix

 \section{BNE under Public Information and Common Value Payoff States}

The expectation of the payoff state $\gamma$ given two Gaussian signals (prior $\mu$ and public signal $\bar \omega$) as follows
\begin{equation}\label{eq_exp_common}
E[\gamma|\omega_{i}=\bar{\omega}] = (1- \xi_{i})\mu + \xi_{i} \bar{\omega}
\end{equation}
where 
$\xi_{i}(\nu) = \frac{var(\gamma)}{var(\gamma) + var(\bar \omega)},$
and $\nu$ is the covariance matrix of the distribution $\zeta(\omega | \gamma)$.


\begin{lemma}\label{thm_public}
Bayesian Nash equilibrium of LQG network game given public signals $\bar \omega$ and common payoff state $\gamma$ can be represented by the following function
\begin{equation}\label{eq_public_common}
a^{*}_{i}(\bar{\omega}) = E[\gamma | \bar{\omega}] [H^{-1} \boldsymbol{1}]_{i} \qquad \forall i \in \ccalN,
\end{equation}
where $\bbone\in \mathbb{R}^n$ is a vector of ones, and $[\cdot]_{i}$ indicates the $i$th element of a vector.
\end{lemma}
\begin{proof}
First order condition of the expectation of the utility function in \eqref{utility} with respect to $a_i$ yields 
\begin{align}\label{eq_first_order}
\frac{\partial E[u_{i}|\{\omega_{i}=\bar{\omega}\}] }{\partial a_{i}} &= -H_{ii}a^{*}_{i}(\bar{\omega}) - \sum_{i\neq j}H_{ij} E[a^{*}_{j}|\{\omega_{i}=\bar{\omega}\}]\nonumber\\ &+ E[\gamma|\{\omega_{i}=\bar{\omega}\}]  =0, \forall i \in \ccalN
\end{align}
We incorporate \eqref{eq_exp_common} into \eqref{eq_first_order}:
\begin{align}\label{eq_first_order2}
 H_{ii}a^{*}_{i}(\bar{\omega}) = - \sum_{i\neq j}H_{ij} E[a^{*}_{j}|\bar{\omega}]+ (1- \xi_{i})\mu + \xi_{i} \bar{\omega} =0, \forall i \in \ccalN
\end{align}
We assume agent $i \in \ccalN$'s strategy is linear in its information $a^{*}_{i}(\bar{\omega}) = \alpha_{i1}\bar{\omega}+  \alpha_{i2}\mu$ with coefficients $\alpha_{i1}$ and $\alpha_{i2}$.  We substitute linear actions in \eqref{eq_first_order2} to get
\begin{align}
 H_{ii}(\alpha_{i1}\bar{\omega}+  \alpha_{i2}\mu) 
 &=  - \sum_{i\neq j}H_{ij}(  \alpha_{j1}\bar{\omega}+  \alpha_{j2}\mu) \nonumber\\&+ (1- \xi_{i})\mu + \xi_{i} \bar{\omega}   =0, \forall i \in \ccalN \label{eq_first_order_actions2}
\end{align}
%
We solve for the action coefficients $\bbalpha_{1}=[\alpha_{11},\dots,\alpha_{n1}]\in \mathbb{R}^n$ and $\bbalpha_{2}=[\alpha_{12},\dots,\alpha_{n2}]\in \mathbb{R}^n:$ $\bbalpha_{1}=\boldsymbol{1}-\bbalpha_{2}  = H^{-1}\xi$
where $\xi = [\xi_1,\dots,\xi_n]$ and $\xi_i$ is as in \eqref{eq_exp_common}. Thus, $a^{*}(\bar{\omega}) =  H^{-1}\xi \bbone\bar{\omega} + (I-H^{-1})\xi \bbone\mu$ where $I$ is the identity matrix. \eqref{eq_public_common} follows from rearranging terms in $a^{*}$ and using \eqref{eq_exp_common}.
\end{proof}

\end{document}